\documentclass[aps,prl,preprint,floats,twocolumn,floatfix,english]{revtex4}
\usepackage[T1]{fontenc}
\usepackage[latin1]{inputenc}
\usepackage{amsmath}
\usepackage{setspace}
\usepackage{epsfig}

\makeatletter
\providecommand{\LyX}{L\kern-.1667em\lower.25em\hbox{Y}\kern-.125emX\@}

\usepackage{babel}
\makeatother
\begin{document}
\title{Spatial snowdrift game with myopic agents}

\author{Marko Sysi-Aho $^1$ }
\email[]{msysiaho@lce.hut.fi}
\author{Jari Saram\"{a}ki $^1$ }
\author{J\'{a}nos Kert\'{e}sz $^{1,2}$ }
\author{Kimmo Kaski $^{1}$ }

\affiliation{$^1$Laboratory of Computational Engineering, Helsinki
University of Technology, Espoo, Finland\\
$^2$Department of Theoretical Physics,
         Budapest University of Technology and Economics, Budapest, Hungary}

\begin{abstract}
We have studied a spatially extended 
snowdrift game, in which the players are located on the sites of 
two-dimensional square lattices and repeatedly have to choose one of the two strategies, either cooperation (C) or defection (D). 
A player interacts with its nearest neighbors only, and 
aims at playing a strategy which maximizes its instant pay-off, 
assuming that the neighboring agents retain their strategies. 
If a player is not content with its current strategy, it will change
it to the opposite one with probability $p$ next round. Here we show through simulations and analytical 
approach that these rules result in cooperation levels, which differ 
to large extent from those obtained using the replicator dynamics.

\end{abstract}

\maketitle

\section{Introduction}
\label{intro}
Understanding the emergence and persistence of cooperation is one of 
the central problems in evolutionary biology and socioeconomics  
\cite{MaynardSmith1995,Neumann1944}. In investigating this problem 
the standard framework utilized is evolutionary game theory  
\cite{Neumann1944,MaynardSmith1973,fudenberg}. Especially two models, the 
Prisoner's Dilemma \cite{Rapoport1965,Axelrod1981,Axelrod1988} and its  
variation, the snowdrift game \cite{MaynardSmith1973,Sugden1986}, 
have attracted most attention. In both games, the players can either 
cooperate for common good, or defect and exploit other players in 
attempt to gain benefits individually. In the Prisoner's Dilemma, 
the precondition is that it pays off  to be non-cooperative. 
Because of this, defection is the only evolutionarily stable 
strategy (ESS) in populations which are fully mixed, i.e. where each  
player interacts with any other player \cite{Smith1982}. However, 
several models which are extensions of the Prisoner's Dilemma have 
proved to sustain cooperation. These models include those in which 
the players are assumed to have memory of the previous interactions 
\cite{Axelrod1984}, or characteristics that allow cooperators and 
defectors to distinguish each other \cite{Epstein1998}, or players 
are spatially distributed \cite{hauert,Lindgren1997,Nowak1992}.  
 
A typical spatial game is such where player-player interactions only 
take place within restricted neighborhoods on regular lattices 
\cite{Nowak1992,Doebeli1998,Szabo1998,Szabo2000} or on complex 
networks \cite{Zimmermann2004}. These games have been found to 
generate highly complex behavior and enable the persistence of 
cooperation. Regarding the latter, the opposite was recently seen 
in the case of the snowdrift game played on a two-dimensional 
lattice \cite{hauert}, where the spatial structure resulted in 
decreased cooperator densities compared to the fully mixed  
``mean-field'' case. This result was surprising, as intermediate 
levels of cooperation persist in unstructured snowdrift games, 
and the common belief has been that spatial structure is usually 
beneficial for sustained levels of cooperation. 
 
In these studies the viewpoint has largely been that of biological 
evolution, as represented by the so-called \emph{replicator dynamics}  
\cite{fudenberg,Hofbauer1998,Nowak2004}, where the fraction of players 
who use high-payoff-strategies grow (stochastically) in the population 
proportionally to the payoffs. This mechanism can be viewed as 
depicting Darwinian evolution, where the fittest have the largest 
chance of survival and reproduction. Overall, the factors influencing 
the outcomes of these spatially structured games are (i) the rules 
determining the payoffs (e.g. Ref.~\cite{fort2003}), (ii) the topology 
of the spatial structure (e.g. Ref.~\cite{Szabo2000}), and  (iii) the 
rules determining the evolution of each player's strategy (e.g. 
Ref.~\cite{Meyer1999,Traulsen2004}). We have studied the effect of 
changing  the strategy evolution rules (iii) in the two-dimensional 
snowdrift game similar  to that discussed in Ref.~\cite{hauert}. 
In our version, the rules have been defined in such a way that changes 
in the players' strategies represent player \emph{decisions} instead 
of different strategy genotypes in the next evolutionary generation 
of players. Thus, the time scale of the population dynamics in our  
model can be viewed to be much shorter than evolutionary time scales.  
Instead of utilizing the evolution-inspired replicator dynamics, 
we have endowed the players with primitive ``intelligence'' in the 
form of local decision-making rules determining their strategies. 
We show with simulations and analytic approach that these rules 
result in cooperation levels which differ largely from those 
obtained using the replicator dynamics.  

In this study we will concentrate on an adaptive snowdrift game, 
with agents interacting with their nearest neighbor agents on a 
two-dimensional square lattice. In what follows we first describe 
our spatial snowdrift model and then analyze its equilibrium states. 
Next we present our simulation results and finally draw some 
conclusions. 
 
\section{Spatial Snowdrift Model} 
\label{sec:2}  
The snowdrift model\footnote{Commonly known as hawk-dove or chicken
  game also.} can be illustrated with a situation in which 
two cars are caught in a blizzard and there is a snowdrift blocking 
their way. The cars are equipped with shovels, and the drivers have  
two choices: either start shoveling the road open or remain in the 
car. If the road is cleared, both drivers gain the benefit $b$ of 
getting home. On the other hand, clearing the road requires some work, 
and cost $c$ can be assigned to it ($b>c>0$). If both drivers are 
cooperative and willing to shovel, this workload is shared between 
them, and both of them gain total benefit of $R=c-b/2$. If both choose 
to defect, i.e. remain in their cars, neither one gets home and 
thus both obtain zero benefit $P=0$. If only one of the drivers shovels, 
both get home, but the defector avoids the cost and gains benefit $T=b$, 
whereas the cooperator's benefit is reduced by the workload, i.e. $S=b-c$.  
 
The above described situation can be presented with the bi-matrix 
\cite{gibbons} (Table \ref{payoffs}), where   
\begin{equation} 
T > R > S > P. 
\label{rank} 
\end{equation}
In case of the so called one-shot game, each player has two available 
strategies, namely defect (D) or cooperate (C). The players choose their 
strategies simultaneously, and their individual payoffs are given by 
the appropriate cell of the bi-matrix. By convention, the payoff  
to the so-called row player is the first payoff given, followed by the payoff  
of the column player. Thus, if for example player 1 chooses D and player 2 
chooses C, then player 1 receives the payoff T and player 2 the payoff S. 
 
The best action depends on the action of the co-player such that defect if 
the other player cooperates and cooperate if the other defects. A simple 
analysis shows that the game does not have \emph{stable evolutionary strategy} 
\cite{Hofbauer1998}, if the agents use only pure strategies, i.e., 
they can choose either to cooperate or to defect with probability 
one, but they are not allowed to use a strategy which mixes either 
of these actions with some probability $q \in (0,1)$. This leads to 
stable existence of cooperators and defectors in well-mixed 
populations \cite{hauert}. 

\begin{table} 
\caption{Snowdrift game. Player 1 chooses an action from the rows and 
  player 2 from the columns. By convention, the payoff to the row 
  player is the first payoff given, followed by the payoff of the column 
  player.}
\label{payoffs} 
\begin{center} 
\begin{tabular}{lll}
\hline\noalign{\smallskip} 
  & D   & C     \\ 
\noalign{\smallskip}\hline\noalign{\smallskip}
D & P, P & T, S \\ 
C & S, T & R, R \\ 
\noalign{\smallskip}\hline
\end{tabular} 
%\vspace*{5cm}
\end{center} 
\end{table} 
 
In order to study the effect of spatial structure on the snowdrift game,  
we set the players on a regular two-dimensional square lattice 
consisting of $m$ cells. We adopt the notation of Ref.~(\cite{schweitzer}) 
and identify each cell by an index $i=1,\ldots,m$ 
which also refers to its spatial position. Each cell, representing a player, 
is characterized by its strategy $s_i$, which can be either to cooperate  
($s_i=1$) or to defect ($s_i=0$). The spatio-temporal distribution of the  
players is then described by $S=(s_1,\ldots,s_m)$ which is an element of 
a $2^m$ dimensional hypercube. Then every player -- henceforth called an 
\emph{agent} -- interacts with their $n$ nearest neighbors. We use either 
the Moore neighborhood in which case each agent has $n=8$ neighbors, in 
N,NE,E,SE,S,SW,W and NW, or the von Neumann neighborhood in which case 
each agent has $n=4$ neighbors, in N,E,S and W compass directions \cite{adachi}. 
We require that an agent plays \emph{simultaneously} with all its $n$ 
neighbors, and define the payoffs for this $(n+1)-player$ game  
such that an agent $i$ who interacts with $n_c^i$ 
cooperators and $n_d^i$ defectors, $n_c^i+n_d^i=n$, gains a benefit of 
\begin{eqnarray} 
u_i(s_i=0) & = & n_c^iT + n_d^iP \label{neiguta}\\ 
u_i(s_i=1) & = & n_c^iR + n_d^iS,  
\label{neigutb} 
\end{eqnarray}  
from defecting or cooperating, respectively.  
 
For determining their strategies, the agents are endowed with primitive  
decision-making capabilities. The agents retain no memory of the past, and  
are not able to predict how the strategies of the neighboring agents will  
change. Every agent simply assumes that the strategies of other agents  
within its neighborhood remain fixed, and chooses an action that maximizes  
its own payoff. In this sense the agents are myopic. The payoff is maximized, if an agent (a) defects when  
$u_i(0) > u_i(1)$, and (b) cooperates when $u_i(1) > u_i(0)$. If (c)  
$u_i(0) = u_i(1)$ the situation is indifferent. Using
Eqs.~(\ref{neiguta}) and (\ref{neigutb})  
we can connect the preferable choice of an agent and the payoffs of the game.  
Let us denote  
\begin{equation} 
\frac{1}{r}=1+\frac{S-P}{T-R}. 
\label{cbr} 
\end{equation} 
Then, if 
\begin{eqnarray} 
\frac{n_c^i}{n} & > & 1-r \textrm{ defecting is 
  profitable, or if} \label{boundarya}\\ 
\frac{n_c^i}{n} & < & 1-r \label{boundaryb}\textrm{ cooperating is 
  profitable, or if} \\ 
\frac{n_c^i}{n} & = & 1-r \textrm{ choices are indifferent}. 
\label{boundaryc} 
\end{eqnarray}  
 
Thus, for each individual agent, the ratio $r$ determines a following 
decision-boundary 
\begin{equation} 
\theta = n(1-r), \label{theta} 
\end{equation} 
which depends on the neighborhood size $n$ and the ``temptation''  
parameter $r$. Because $r$ is determined only by the differences  
$T-R$ and $S-P$, we can fix two of the payoff values, say $R=1$ and $P=0$.  
Based on the above, we define the following rules for the agents: 
\begin{enumerate} 
\item If an agent $i$ plays at time $t$ a strategy $s_i(t) \in \{0,1\}$ for 
  which $u_i(s_i) \geq u_i(1-s_i)$, then at time $t+1$ the agent plays 
$s_i(t+1) =  s_i(t)$. 
\item  If an agent $i$ plays at time $t$ a strategy $s_i(t) \in \{0,1\}$ for 
  which $u_i(s_i) < u_i(1-s_i)$, then at time $t+1$ the agent plays  
$s_i(t+1) = 1  - s_i(t)$ with probability $p$, and $s_i(t+1)=s_i(t)$  
with probability $1-p$. 
\end{enumerate} 
Hence, the strategy evolution of an individual agent is determined by  
the current strategies of the other agents within its neighborhood, with 
the parameter $p$ acting as a ``regulator'' which moderates the rate 
of changes.

\section{Equilibrium states} 
\label{sec:3}
A spatial game is in stable state or equilibrium if retaining the current 
strategy is beneficial for all the agents~\cite{fudenberg}. There can be  
numerous equilibrium configurations, depending on the temptation  
parameter $r$, geometry and size of the $n$-neighborhood, 
and the size and boundary conditions of the lattice upon which the 
game is played. An aggregate quantity of particular interest is the 
fraction of cooperators $F_c$ in the whole population (or, equivalently, 
that of the defectors $F_d$). Below, we derive limits for $F_c$, first 
in a ``mean-field'' picture based cooperator densities within neighborhoods 
and then by investigating local neighborhood configurations. 
 
\subsection{Mean-field limits for cooperator density} 
\label{sec:3a} 
Without detailed knowledge of local equilibrium configurations we can 
already derive some limits for the fraction of cooperators in equilibrium. 
Let us consider a square lattice with $m=L\times L$ cells with periodic 
boundary conditions, where $L$ is the linear size of the lattice, and 
assume that $k$ cells are occupied by cooperators. We denote by  
$a_j$ the number of those agents who have $j$ cooperators each in their 
$n$-neighborhood, excluding the agents themselves, and denote the local 
density of cooperators in such neighborhoods by $f_c=j/n$. Hence, the 
total amount of cooperators $k$ can be written in terms of the 
densities as follows 
 
\begin{equation} 
k = \sum_{j=0}^n a_j f_c = \sum_{j=0}^n a_j \frac{j}{n}. 
\label{id1} 
\end{equation} 
 
From Eqs.~(\ref{boundarya})-(\ref{boundaryc}) we can infer that a cooperator will retain 
its current strategy, if it has at most $c$ cooperators in its 
$n$-neighborhood, where $c$ is the integer part of $\theta=n(1-r)$. 
Similarly, a defector will remain a defector if it has more than $c$ 
cooperators in its neighborhood. Thus, in equilibrium, all agents 
having $j\leq c$ cooperators in their neighborhood are likewise 
cooperators, and thus $\sum_{j=0}^c a_j = k$. We denote by 
$\left<f_{c|c}\right> = \frac{1}{k}\sum_{j=0}^c a_j \frac{j}{n}$ 
the average density of cooperators as the nearest neighbors of 
cooperators. Similarly, $\left<f_{c|d}\right>$ denotes the average 
density of cooperators as the nearest neighbors of defectors, i.e. 
$\left<f_{c|d}\right> = \frac{1}{m-k}\sum_{j=c+1}^n a_j \frac{j}{n}$.  
Then we can write Eq.~(\ref{id1}) as 
\begin{equation} 
k=k\left<f_{c|c}\right> + (m-k)\left<f_{c|d}\right>. 
\label{id2} 
\end{equation} 
 
\begin{figure} 
\begin{center}
\resizebox{0.4\textwidth}{!}{
\includegraphics{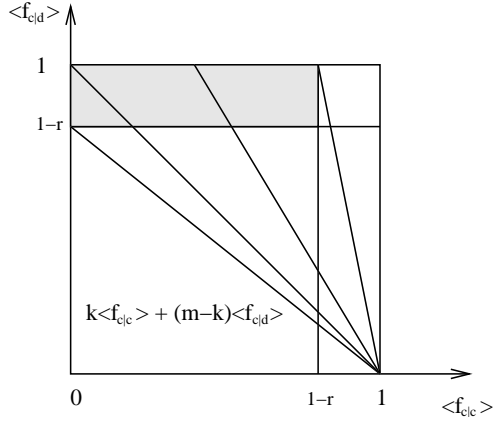}} 
\caption{In equilibrium the average density of cooperators in the 
  nearest neighborhood of defectors must be $1-r \leq \left < f_{c|d} \right > \leq 1$ and 
  in the nearest neighborhood of cooperators $0 \leq \left < f_{c|c} \right > \leq 1-r$ 
  (shaded area). If the total number of players in the lattice is $m$, 
  the lines $k\left < f_{c|c} \right > + (m-k)\left < f_{c|d} \right > 
  = k$ depict the identity of $k$ cooperators in the lattice. Equilibrium is 
  not possible, if the fraction of cooperators $F_c = k/m$ is such that the lines 
  do not pass through the shaded area.\label{fig1}} 
\end{center} 
\end{figure} 
 
The density  $f_{c|c}$ of cooperators around each cooperator is bounded: 
$f_{c|c}\geq 0$, $f_{c|c}\leq c/n$, and as $c \leq \theta = n(1-r)$, the 
relation $0 \leq \left<f_{c|c}\right> \leq 1-r$ holds for the average density. 
Similarly, the density of cooperators around each defector $f_{c|d}$ can be 
at most $1$ and is at least $(1-r)$, and thus the average density 
$1-r \leq \left<f_{c|d}\right> \leq 1$. Using these relations together 
with Eq.~(\ref{id2}) we obtain the following limits for the density of 
cooperators $F_c=k/m$ in the whole agent population 
(see also Fig.~\ref{fig1}): 
\begin{equation} 
\frac{1-r}{2-r} \leq F_c \leq \frac{1}{r+1}. 
\label{loup} 
\end{equation} 
 
\subsection{Local equilibrium configurations} 
 \label{sec:3b}
In the above derivation we ignore how the strategies can actually be 
distributed in the lattice. Hence, it is of interest to examine possible 
local equilibrium configurations of the player strategies. Again, 
Eqs.~(\ref{boundarya})-(\ref{boundaryc}) tell us how many cooperative neighbors each 
defector or cooperator can have in the equilibrium state. The number 
of cooperators around each agent depends on the value of the temptation 
parameter $r$, and for a given value of $r$ the lattice has to be filled 
such that these conditions hold for the neighborhood of each agent. 
In a lattice with periodic boundary conditions, the lattice size 
$m=L_X \times L_Y$ and the neighborhood size $n$ obviously have an 
effect on the elementary configurations. Hence, we restrict ourselves 
to infinite-sized lattices, filled by repeating elementary configuration 
blocks, and look for the resulting limits on the cooperator density $F_c$.  
Note that these conclusions also hold for finite lattices with periodic 
boundary conditions, if $L_X$ and $L_Y$ are integer multiples of $X$ 
and $Y$, respectively, where $X\times Y$ is the elementary block size. 
Here, we will restrict the analysis to the case of the Moore neighborhood 
with $n=8$. 
 
\begin{table} 
\caption{ Limits for the equilibrium fraction of cooperators based on 
        repeating elementary configuration blocks. When $r_l < r < r_u$, 
  the number of cooperators in each defector's neighborhood $N_{c|d}$  
  must be at least $9-i$ and the number of cooperators in each 
  cooperator's neighborhood $N_{c|c}$ at most $8-i$. Considering 
  possible repeating configuration blocks which fulfill these 
  conditions, we obtain lower limits $F_{c,L}$ and upper limits 
  $F_{c,U}$ for the density of cooperators.} 
\label{tab1} 
\begin{center} 
\begin{tabular}{lllccll}
 \hline\noalign{\smallskip}
i &$r_l$  & $r_u$ & $N_{c|d} \ge$ & $N_{c|c} \le$ & $F_{c,L}$ &
$F_{c,U}$ \\ 
\noalign{\smallskip}\hline\noalign{\smallskip} 
1&$0$   & $1/8$ & 8  & 7  & $3/4$  & $8/9$ \\ 
2&$1/8$ & $2/8$ & 7  & 6  & $2/3$  & $4/5$ \\ 
3&$2/8$ & $3/8$ & 6  & 5  & $1/2$  & $2/3$ \\ 
4&$3/8$ & $4/8$ & 5  & 4  & $1/2$  & $2/3$ \\ 
5&$4/8$ & $5/8$ & 4  & 3  & $4/9$  & $1/2$ \\ 
6&$5/8$ & $6/8$ & 3  & 2  & $1/3$  & $1/2$ \\ 
7&$6/8$ & $7/8$ & 2  & 1  & $2/9$  & $1/3$ \\ 
8&$7/8$ & $8/8$ & 1  & 0  & $1/9$  & $1/4$ \\ 
\noalign{\smallskip}\hline
\end{tabular} 
\end{center} 
\end{table} 
 
\begin{figure} 
\begin{center} 
\resizebox{0.4\textwidth}{!}{
\includegraphics{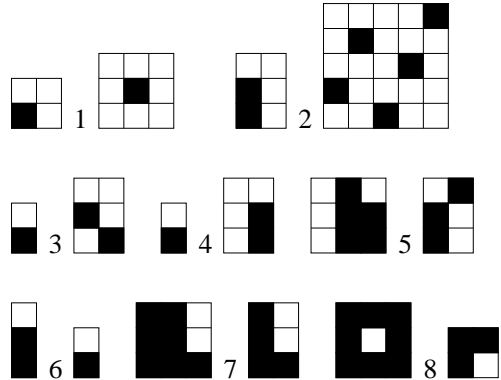}} 
\caption{ Examples of elementary configuration blocks which can be 
  repeated without overlap to fill an infinite lattice, for various  
  values of $r$. The numbering refers to $i$ in Table \ref{tab1}. A 
black cell denotes a defector while an empty cell denotes a cooperator. 
For a particular number the lower limit of density is obtained by filling 
the lattice with the blocks on the left, and the upper by using the 
blocks on the right. \label{fig2}} 
\end{center} 
\end{figure} 
 
As an example, consider the local configurations when $r=0.1$,  
and hence the decision boundary value $\theta = n(1-r) = 7.2$.  
Thus, from Eqs.~(\ref{boundarya})-(\ref{boundaryc}) one can infer that in equilibrium 
all defectors should have more than $7.2$ cooperators in their 
Moore neighborhoods. Because the number of cooperating neighbors 
can take only integer values, this means that every one of the $n=8$ 
neighbors of a defector should be a cooperator. On the other hand, 
from Eqs.~(\ref{boundarya})-(\ref{boundaryc}) we see that the density $f_{c|c}$ of 
cooperators around each cooperator should be less than $1-r$, i.e. 
they should have at most $c=7$ cooperators in their Moore neighborhood. 
The smallest repeated elementary block fulfilling both conditions 
is a $2\times 2$-square with one defector -- when the lattice  
is filled with these blocks, the cooperator density equals $F_c=3/4$ 
(see Fig.~(\ref{fig2}), case 1, left block). On the other hand, both 
requirements are likewise fulfilled with a repeated $3\times 3$-square, 
where the central cell is a defector and the rest are cooperators, 
resulting in the cooperator density of $F_c=8/9$. This configuration 
is illustrated in Fig.~(\ref{fig2}), as case 1, right block. 
   
By continuing the analysis of elementary configuration blocks in similar 
fashion for different values of $r$, we obtain lower and upper limits 
for the fraction of cooperators, which are listed in Table \ref{tab1}.  
The corresponding elementary configuration blocks are depicted 
in Fig.~(\ref{fig2}). The table is read so that when the value of 
the temptation parameter is within the interval $r_l < r < r_u$,  
the number of cooperators in each defector's neighborhood $N_{c|d}$ 
must be at least $9-i$ and the number of cooperators in each 
cooperator's neighborhood $N_{c|c}$ can be at most $8-i$. Here
$r_l=(i-1)/8$, $r_u=i/8$ and $i=1,\ldots,8$ These conditions are 
those of Eqs.~(\ref{boundarya})-(\ref{boundaryc}) and they are fulfilled by the 
configuration blocks depicted in Fig.~(\ref{fig2}), for which the 
minimum and maximum densities of cooperators are $F_{c,L}$ and $F_{c,U}$. 
 
\section{Simulation results} 
 \label{sec:4}
We have studied the above described spatial snowdrift model with 
discrete time-step simulations on a $m=100 \times 100$-lattice with 
periodic boundary conditions. We have specifically analyzed the behavior 
of the cooperator density $F_c$, and equilibrium lattice configurations. 
In the simulations, the lattice is initialized randomly so that each  
cell contains a cooperator or defector with equal probability. However, 
biasing the initial densities toward cooperators or defectors was found to  
have no considerable effect on the outcome of the game. We have simulated 
the game using both the Moore and the von Neumann neighborhoods 
with $n=8$ and $n=4$ nearest neighbors, respectively. In the simulations 
we update strategies of the agents asynchronously \cite{adachi} with the 
random sequential update scheme, so that during one simulation round, 
every agent's strategies are updated in random order. In the following, 
the time scale is defined in terms of these simulation rounds. 
 
\begin{figure} 
\begin{center}
\resizebox{0.4\textwidth}{!}{
\includegraphics{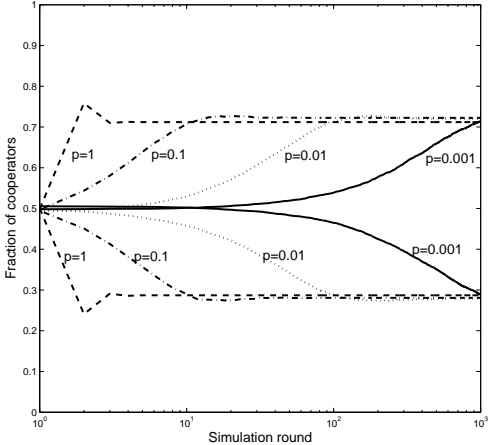}} 
\caption{Dynamics of the fraction of cooperators $F_c$. The upper curves  
that converge to $F_c \sim 0.7$ are for $r=0.2$, and the lower curves
that converge to $F_c \sim 0.3$ are for $r=0.8$. In both cases the probability of 
being discontent is varied as $p=1,0.1,0.01,0.001$ 
from left to right, and the lattice size is $m=100x100$.} 
\label{fig3} 
\end{center} 
\end{figure}

First, we have studied the development of the cooperator density  
$F_c$ as a function of time. As expected, the probability 
$p$ of discontent agents changing their strategies plays the role  
of defining the convergence time scale only\footnote{The 
role of $p$ would be more important if synchronous update rules were 
used. In that case $p=1$ corresponds to a situation where each 
discontent agent simultaneously changes its strategy to  
the opposite. This, then, could result in a frustrated situation 
with oscillating cooperator density. However, 
small enough values of $p$ should damp these oscillations, 
resulting in static equilibrium.}, as in the long run $F_c$ converges 
to a stable value irrespective of $p$. This is depicted in Fig.~\ref{fig3}, 
which shows $F_c$ as function of time for several values of $p$ and 
two different values of the temptation $r$. In these runs, we have used 
the Moore neighborhood, i.e. $n=8$. In all the studied cases, $F_c$ 
turns out to converge quite rapidly to a constant value, $F_c\sim 0.7$ 
for $r=0.2$ and $F_c\sim 0.3$ for $r=0.8$. 
 
It should be noted that $F_c$ does not have to converge to  
exactly the same stable value for the same $r$; even if the game 
is considered to be in equilibrium, there can be some variance 
in $F_c$, which is also visible in Fig.~\ref{fig3}. However, 
the value of $F_c$ was found to eventually remain stable during 
individual runs, i.e. no oscillations were detected.  
 
%Fig.~(\ref{fig3}) also points out that if we use the Moore 
%neighbourhood and pick $r$ from two different intervals 
%$i/8<r<(i+1)/8$, i.e., for two different $i = 1,\ldots,7$, then the values of 
%$f_c$ tend to converge to different levels. Partly, the explanation for this 
%is given by Eqs.~(\ref{boundarya})-(\ref{boundaryc}) which show how the number of 
%cooperators in each agent's neighbourhood depends on 
%$r$. We discussed about this dependency already in Equilibrium section 
%when we derived upper and lower limits for $f_c$. We can also 
%check that $f_c$ in Fig.~(\ref{fig3}) falls between the lower and 
%upper limits given in Tab.~(\ref{tab1}). For $r=0.2$, 
%$i=2$ in the above interval. From Tab.~(\ref{tab1}) we see that 
%corresponding to $i=2$ the lower limit $f_l(c)=2/3$ and the upper 
%limit $f_l(c)=4/5$. In Fig.~(\ref{fig3}) $f_c$ converges to values 
%slightly over $0.7$ when $r=0.2$ for all used $p$. This clearly falls 
%between the lower and upper limits. Similarly we can check that for 
%$r=0.8$, $i=7$ in Tab.~(\ref{tab1}), and the lower and upper limits for 
%$f_c$ are $f_l(c)=2/9$ and $f_u(c)=1/3$. In Fig.~(\ref{fig3}) the 
%curves for $r=0.8$ fall between these limits. 
 
\begin{figure} 
\begin{center}
\resizebox{0.4\textwidth}{!}{
\includegraphics[angle=270]{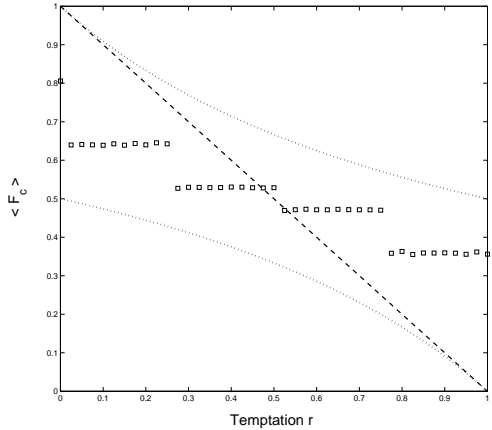}}
\caption{Average fraction of cooperators $\left<F_c\right>$ versus the
  temptation $r$ (squares), 
simulated on a $100\times 100$ lattice with $p=0.1$ using the von
Neumann neighborhood. The values for $\left<F_c\right>$ 
are averages over 500 simulation rounds, where the averaging was started after 500 
initial rounds to guarantee convergence. The dotted lines depict the upper and 
lower limits for $F_c$ of Eq.~(\ref{loup}). The dashed diagonal line is $1-r$.} 
\label{fig4} 
\end{center} 
\end{figure}

\begin{figure*} 
\begin{center}
\resizebox{0.75\textwidth}{!}{
\includegraphics{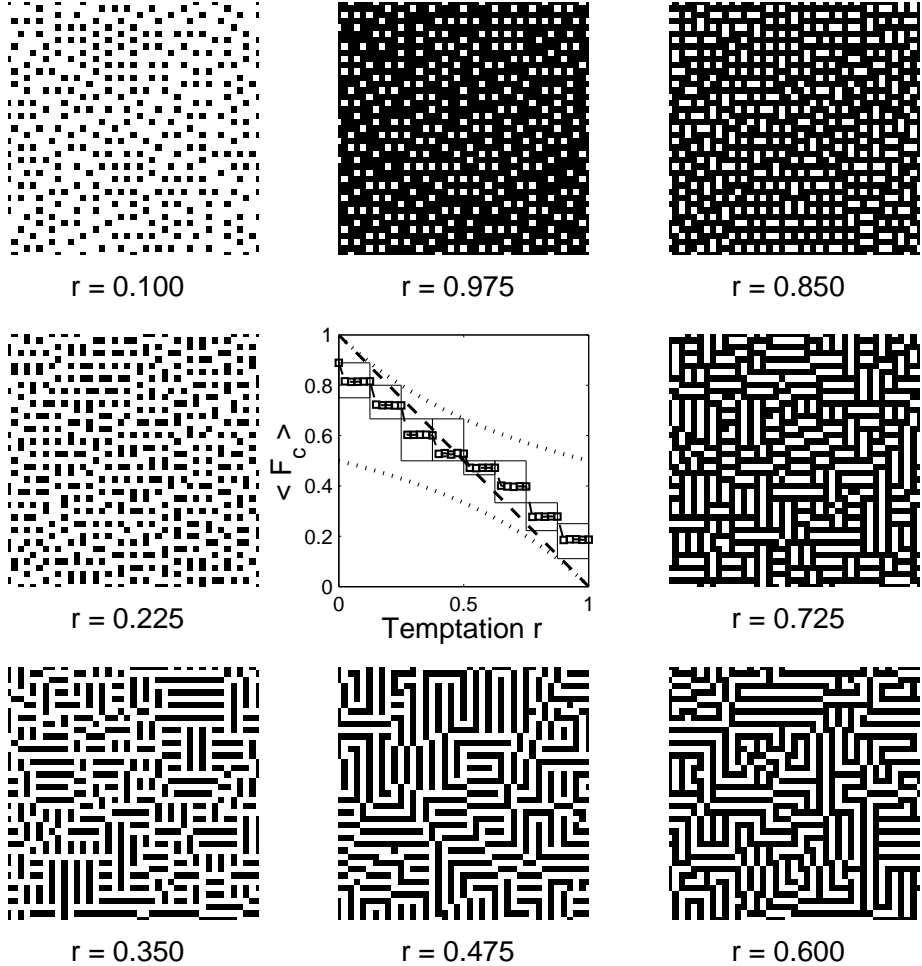}}
\caption{ Example equilibrium configurations of defectors and cooperators on a 
 $m=100\times 100$ lattice for various values of $r$ when the Moore
 neighborhood is used. The configurations 
were recorded after $T=1000$ simulation rounds. Only the middle part of  
the lattice is shown for the sake of clarity. The middle panel depicts the 
average fraction of cooperators $\left<F_c\right>$ in the whole population as a function 
of the temptation $r$ (squares), together with the upper and lower limits of 
Eq.~(\ref{loup}) (dotted lines) and the limits of Table \ref{tab1} (solid
lines). The values of $\left<F_c\right>$ are averages over the last
500 simulation rounds and the dashed diagonal line is $F_c=1-r$, corresponding 
to the fraction of cooperators in the fully mixed case~\cite{fudenberg,hauert,Hofbauer1998}.} 
\label{fig5} 
\end{center} 
\end{figure*}

Next, we have studied the average equilibrium fraction of cooperators 
$\left< F_c\right >$ in the agent population as function of the temptation 
parameter $r$. We let the simulations run for 500 rounds (with $p=0.1$), and 
averaged the fraction of cooperators for the subsequent 500 rounds. 
In all cases, the fraction had already converged before the averaging rounds. 
Fig.~(\ref{fig4}) shows the results for the von Neumann neighborhood ($n=4$), illustrated as
the squares. The dotted lines indicate the upper and lower limits of  
Eq.~(\ref{loup}), and the dashed diagonal line is $F_c=1-r$, corresponding 
to the fraction of cooperators in the fully mixed 
case~\cite{fudenberg,hauert,Hofbauer1998}. The fraction of cooperators 
$\left<F_c\right>$ is seen to follow a stepped curve, with steps corresponding 
to $r=i/n$, where $i=0,\ldots,n$. This is a natural consequence of 
Eqs.~(\ref{boundarya})-(\ref{boundaryc}), where the decision boundary $\theta=n(1-r)$ 
can take only discrete values. A similar picture is given for the Moore neighborhood 
($n=8$) in the middle panel of Fig.~(\ref{fig5}). Furthermore, in the
middle panel of Fig.~(\ref{fig5}) the values 
of $F_c$ fall between the limits given in Table \ref{tab1} for all $r$
as shown with solid lines. 
  
In both cases (i.e. with Moore and von Neumann neighborhoods) cooperation is 
seen to persist during the whole range $r=[0,1]$. This result differs 
largely from the $F_c(r)$-curves of the spatial snowdrift game with 
replicator dynamics~\cite{hauert}, where the fraction of cooperators 
vanished at some critical $r_c$. Hence, we argue that no conclusions 
on the effect of spatiality on the snowdrift game can be drawn without 
taking into consideration the strategy evolution mechanism; local 
decision-making in a restricted neighborhood yields results which are 
different from those resulting from the evolutionary replicator dynamics.   
 
We have also studied the equilibrium lattice configurations for  
various values of $r$. Fig.~(\ref{fig5}) depicts the central part of the  
$100\times 100$-lattice after 1000 simulation rounds using the Moore  
neighborhood and $p=0.1$, with white pixels corresponding to cooperators and 
black pixels to defectors. The values of $r$ have been selected so that 
the equilibrium situation corresponds to each plateau of
$\left<F_c\right>$ illustrated in the central panel.
 
The observed configurations are rather polymorphic, and repeating elementary 
patterns like those in Fig.~(\ref{fig2}) are not seen. This reflects the fact 
that the local equilibrium conditions can be satisfied by various configurations; 
the random initial configuration and the asynchronous update then lead to 
irregular-looking equilibrium patterns, which vary between simulation runs. 
The patterns seem to be most irregular when $r$ is around 0.5; this is because 
then the equilibrium numbers of cooperators and defectors are close to each other, 
and the ways to assign strategies within local neighborhoods are most numerous. 
To be more exact, there are $8 \choose i$ ways to distribute $i$ cooperators in 
the $8$-neighborhood, and if e.g. $3/8<r<4/8$, $i$ is at least $4$ and at most $5$, 
maximizing the value of the binomial coefficient. Hence, the ways of filling the 
lattice with these neighborhoods in such a way that the equilibrium conditions 
are satisfied everywhere are most numerous as well.

\section{Summary and conclusions} 
 \label{sec:5}
We have presented a variant of the two-dimensional snowdrift game, where the 
strategy evolution is determined by agent decisions based on the strategies 
of other players within its local neighborhood. We have analyzed the lower 
and upper bounds for equilibrium cooperator densities with a mean-field 
approach as well as considering possible lattice-filling elementary configuration 
blocks. We have also shown with simulations that this game converges to 
equilibrium configurations with constant cooperator density depending on 
the payoff parameters, and that these densities fall within the derived limits. 
Furthermore, the strategy configurations in the equilibrium state display 
interesting patterns, especially for intermediate temptation parameter values. 
 
Most interestingly, the equilibrium cooperator densities differ largely from those 
resulting from applying the replicator dynamics~\cite{hauert}. With our strategy 
evolution rules, cooperation persists through the whole temptation parameter 
range. This illustrates that one cannot draw general conclusions on 
the effect of spatiality on the snowdrift game without taking the strategy 
evolution mechanisms into consideration -- this should, in principle, apply 
for other spatial games as well. Care should especially be taken when 
interpreting the results of investigations on such games: the utilized 
strategy evolution mechanism should reflect the system under study. We argue 
that especially when modeling social or economic systems, there is 
no \emph{a priori} reason to assume that generalized conclusions can be 
drawn based on results using the evolution inspired replicator dynamics 
approach, where high-payoff strategies get copied and ``breed'' in proportion 
to their fitness. As we have shown here, local decision-making with limited 
information (neighbor strategies are known payoffs are not) can result 
in different outcome.

% Non-BibTeX users please use

\end{document}